\documentclass[aip,jcp,reprint]{revtex4-2} 

\usepackage{graphicx}
\usepackage{enumitem}

\usepackage{mathtools}
\usepackage{graphicx}   
\usepackage{color}
\usepackage{caption}
\usepackage{subcaption}
\usepackage{comment}
\usepackage{bm}

\usepackage{algorithm}
\usepackage{algpseudocode}
\usepackage{algorithmicx}

\usepackage{array}
\usepackage{booktabs}
\usepackage{multirow}

\usepackage{amsmath}
\usepackage{amsfonts}

\usepackage{booktabs}

\begin{document}
\title{Augmenting Human Expertise in Weighted Ensemble Simulations through Deep Learning based Information Bottleneck}
	
\author{Dedi Wang}
\affiliation{Biophysics Program and Institute for Physical Science and Technology, 
University of Maryland, College Park 20742, USA.}

\author{Pratyush Tiwary\footnote{Corresponding author.}}
 
\email{ptiwary@umd.edu}
\affiliation{Department of Chemistry and Biochemistry and Institute for Physical Science and Technology, University of Maryland, College Park 20742, USA.}
\affiliation{University of Maryland Institute for Health Computing, Bethesda 20852, USA.}

\date{\today}
	
\begin{abstract}
The weighted ensemble (WE) method stands out as a widely used segment-based sampling technique renowned for its rigorous treatment of kinetics. The WE framework typically involves initially mapping the configuration space onto a low-dimensional collective variable (CV) space and then partitioning it into bins. The efficacy of WE simulations heavily depends on the selection of CVs and binning schemes. The recently proposed State Predictive Information Bottleneck (SPIB) method has emerged as a promising tool for automatically constructing CVs from data and guiding enhanced sampling through an iterative manner. In this work, we advance this data-driven pipeline by incorporating prior expert knowledge. Our hybrid approach combines SPIB-learned CVs to enhance sampling in explored regions with expert-based CVs to guide exploration in regions of interest, synergizing the strengths of both methods. Through benchmarking on alanine dipeptide and chignoin systems, we demonstrate that our hybrid approach effectively guides WE simulations to sample states of interest, and reduces run-to-run variances. Moreover, our integration of the SPIB model also enhances the analysis and interpretation of WE simulation data by effectively identifying metastable states and pathways, and offering direct visualization of dynamics.
\end{abstract}

\maketitle
\section{Introduction}

Advances in computational hardware and algorithms have made all-atom molecular dynamics (MD) a key tool for studying chemical and biological systems. By implementing physical laws \textit{in silico}, MD simulations allow for detailed tracking of molecular evolution over large spatiotemporal domains—up to thousands of angstroms with atomic precision and up to milliseconds at femtosecond resolution.\cite{Dror2012,Hollingsworth2018} However, MD simulations have their challenges. Many important processes, such as ligand dissociation or protein conformation changes, occur over timescales ranging from milliseconds to hundreds of seconds and involve very high free energy barriers. MD simulations often struggle to overcome these barriers, making such events exceedingly rare and difficult to capture. Even with the most advanced resources available, simulating these processes can still take months or even years.\cite{Shaw2021}

A variety of methods, generally called enhanced sampling techniques, have emerged as a solution to the rare event problem. For a comprehensive review of the enhanced sampling methods used in molecular dynamics simulations, interested readers can refer to Refs.  \onlinecite{Bernardi2015,Yang2019,Henin2022,Mehdi2024}. In this work, we majorly focus on an increasingly popular segment-based sampling method called weighted ensemble (WE).\cite{huber1996weighted,zuckerman2017weighted} The WE method has demonstrated its versatility across a spectrum of challenges, from protein folding,\cite{adhikari2019computational} protein-protein binding,\cite{saglam2019protein} and protein-ligand unbinding,\cite{lotz2018unbiased} to the large-scale dynamics of the SARS-CoV-2 spike protein.\cite{sztain2021glycan} Motivated by these real-world applications, the field is undergoing rapid evolution, which is further catalyzed by the emergence of easy-to-use and developer-friendly tools such as WESTPA 2.0.\cite{russo2022westpa}

At the heart of the WE framework lies the process of mapping the configuration space onto a low-dimensional collective variable (CV) space, followed by partitioning it into bins. This mapping and binning process enables the dynamic splitting and merging of trajectories based on their progress along the identified CVs, thereby enhancing the sampling of rare events. The selection of appropriate CVs and binning schemes can significantly influence the algorithm's performance. Voronoi polyhedra offer a convenient method to tessellate spaces of any dimensionality. This technique has been leveraged to enable adaptive binning in a hierarchical manner for the WE method.\cite{zhang2010weighted,dickson2014wexplore}
Based on the finite-temperature string method, Ref. \onlinecite{adelman2013simulating} adaptively learns a string of Voronoi cells to restrict the WE sampling to a one-dimensional path embedded in a multi-dimensional space defined by some CVs. Later, a minimal adaptive binning (MAB) scheme\cite{torrillo2021minimal} is proposed to automate the placement of bins along a chosen CV, which effectively surmounts large free energy barriers encountered during the WE simulation. Alternatively, recent advancements propose optimizing both bin construction and trajectory allocation based on a newly developed mathematical formulation.\cite{aristoff2023weighted} Despite these strides towards more automatic binning schemes, the determination of CVs in WE simulations still heavily relies on expert intuition. 

Moreover, as an unbiased method, the WE method suffers from slow convergence. To speed up the convergence, Ref. \onlinecite{ahn2021gaussian} uses Gaussian accelerated molecular dynamics to generate an initial set of configurations.
These configurations, weighted appropriately, are then utilized to seed the WE simulations, enhancing convergence. Similarly, Ref. \onlinecite{ojha2023deepwest} adopts a comparable approach, enhancing the initial state distribution for WE simulations by constructing a Markov state model (MSM) from short unbiased trajectories. Moreover, history-augmented Markov state models (haMSMs) have emerged as potent tools for estimating stationary distributions and rate constants from transient, unconverged WE data.\cite{copperman2020accelerated} These estimates can then also serve as effective starting points for initiating new WE simulations, thereby accelerating the convergence of the simulation. 

In this work, we present a hybrid approach that integrates the recently proposed data-driven State Predictive Information Bottleneck (SPIB) method with expert knowledge to augment the WE simulations. SPIB offers unique capabilities in constructing CVs: It automatically clusters MD conformations into several metastable states using a pre-specified lag time parameter. By appropriately selecting the lag time, SPIB can adaptively adjust the number of metastable states, filtering out states that are short-lived for the specified lag time. Consequently, SPIB can focus on slow state-to-state transitions, effectively encoding all identified states and transitions between them to learn low-dimensional CVs. This makes SPIB particularly suitable for systems with multiple metastable states of interest. At the same time, while SPIB is data-driven and automated, we also recognize the value of expert intuition in differentiating states from one another within complex molecular systems. To fully harness this intuition, we incorporate the usage of expert-based CVs to complement SPIB-learned CVs. The idea is straightforward: employ SPIB-learned CVs to enhance sampling in explored regions, while using expert-based CVs to further establish new regions of interest for possible exploration. This integration combines the strengths of both approaches: expert-based CVs excel in extrapolation, making them effective and reliable at guiding exploration of new regions, while data-driven CVs are adept at capturing arbitrary complex dynamics, which are often difficult for experts to fully anticipate. 

We evaluate the effectiveness of the proposed hybrid algorithm using two model systems, alanine dipeptide and the mini folding protein CLN025, a mutant of Chignolin (PDB ID: 2RVD). Our results underscore the success of this hybrid approach in steering WE simulations towards sampling the targeted states for these systems, concurrently mitigating run-to-run variances in rate estimates. Additionally, the obtained SPIB model offers improved interpretation of the underlying dynamics by identifying crucial metastable states and pathways, as well as providing a 2D visualization of the dynamics. The proposed hybrid SPIB-WE approach is available as an open-source Python package at \href{https://github.com/wangdedi1997/spib_we}{https://github.com/wangdedi1997/spib\_we}. This package serves as a plugin for WESTPA 2.0, facilitating straightforward integration for WE users.

\begin{figure}[ht]
    \centering
    \includegraphics[width=.49\textwidth]{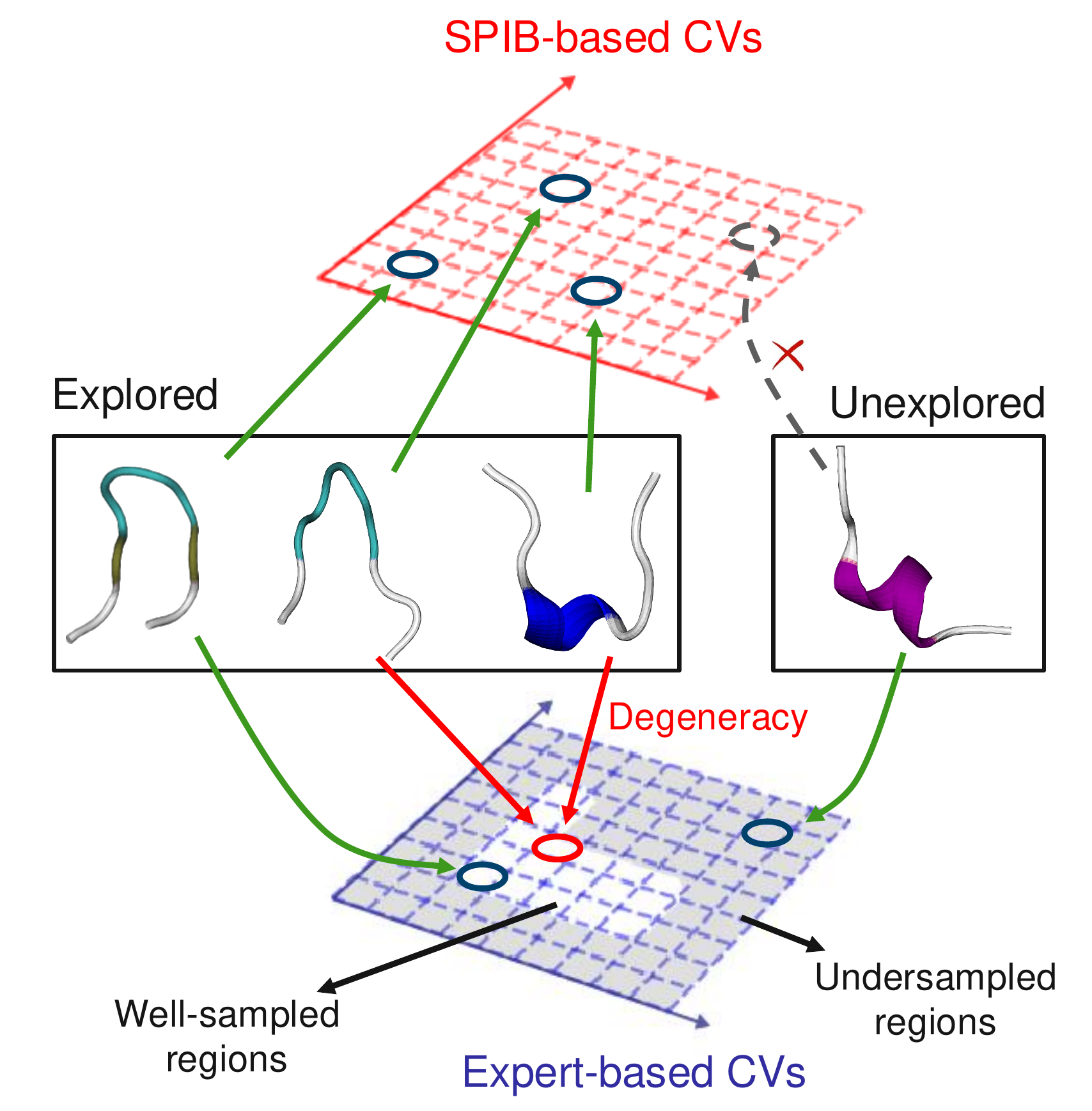}
    \caption{Comparison of SPIB-based CVs and expert-based CVs. Expert-based CVs excel in extrapolation but may struggle with lifting intermediate state degeneracy. In contrast, SPIB-based CVs effectively distinguish distinct states and capture complex dynamics.}
    \label{fig:illustration}
\end{figure}

\begin{figure}[ht]
    \centering
    \includegraphics[width=.4\textwidth]{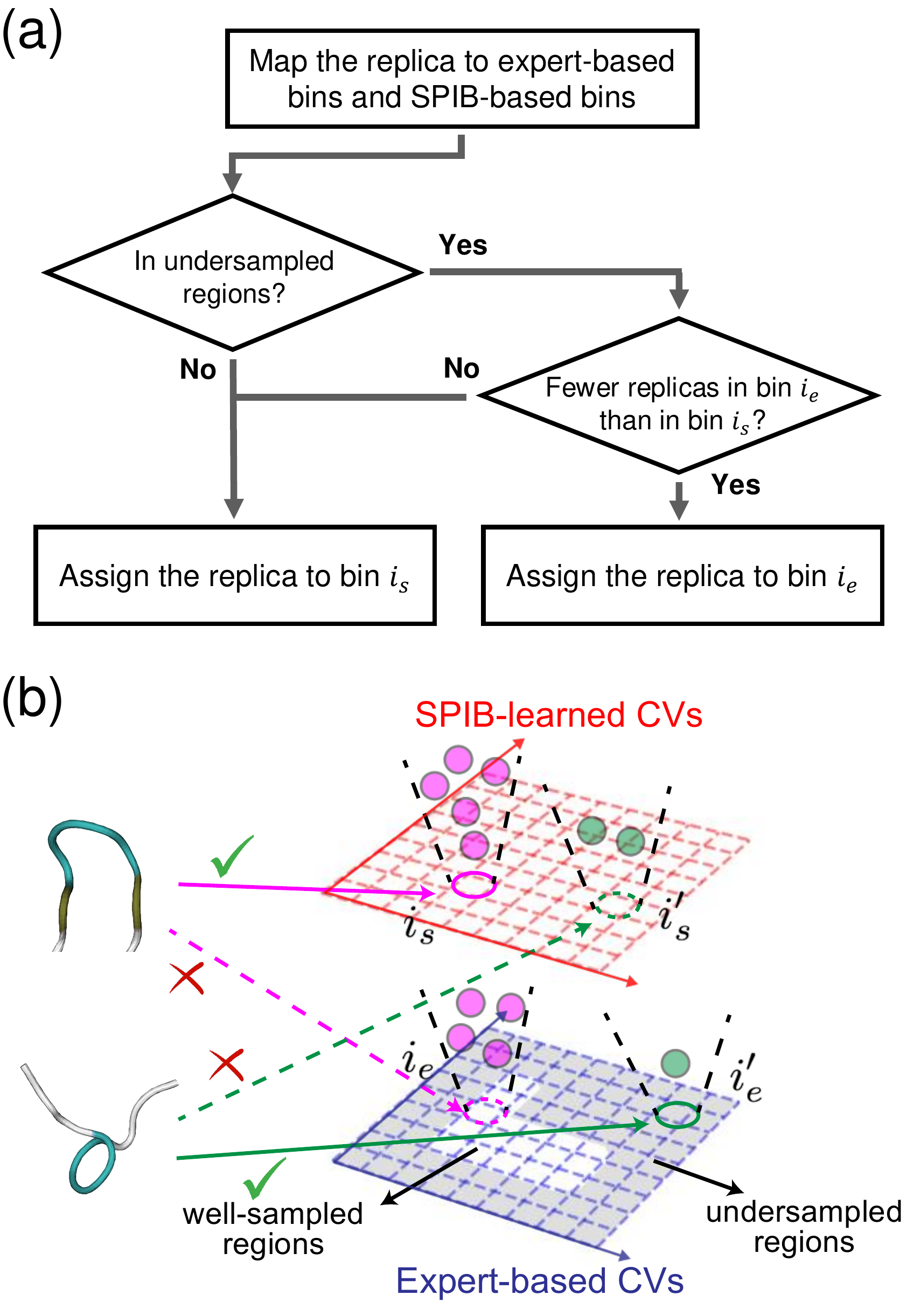}
    \caption{Hybrid approach flowchart and illustration. (a) Algorithm flowchart illustrating how the hybrid approach assigns replicas to bins. (b) Illustration of the hybrid approach combining expert-based CVs and data-driven SPIB-learned CVs for two instances of a protein conformation (top and bottom). Replicas are first mapped both to the expert-based and SPIB-learned CVs, denoted with subscripts $e$ and $s$ respectively. The top (bottom) replica is mapped to bin $i_e$ ($i'_e$) based on expert-based bins and bin $i_s$ ($i'_s$) based on SPIB-learned bins, as indicated by the arrows. Undersampled regions, defined by expert-based bins, are shaded in the bottom right figure. The total number of replicas mapped to these bins is represented by the number of circles. Since bin $i_e$ is not in an undersampled region, the top structure is assigned to bin $i_s$ (solid magenta arrow), regardless of the number of replicas in bins $i_e$ and $i_s$. However, since bin $i'_e$ is in an undersampled region and contains fewer replicas (with only one replica) compared to bin $i'_s$ (which has two replicas), the bottom replica is assigned to $i'_e$ (solid green arrow).}
    \label{fig:alg}
\end{figure}

\begin{figure}[ht]
    \centering
    \includegraphics[width=.4\textwidth]{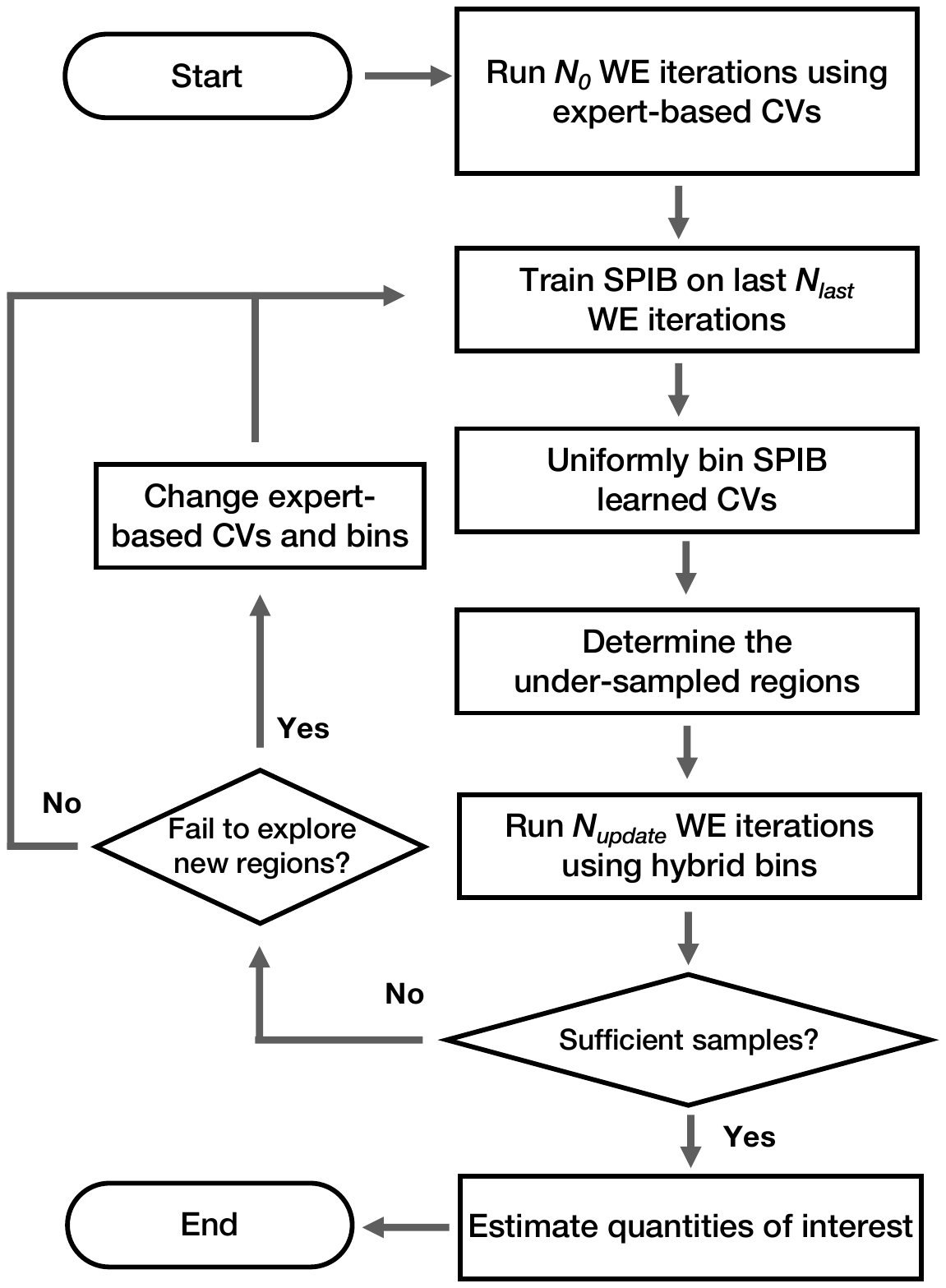}
    \caption{
    Workflow diagram illustrating the hybrid SPIB-WE protocol. First, $N_0$ iterations of WE simulations are conducted using expert-based CVs. Next, SPIB is trained on the collected data to construct low-dimensional CVs for guiding the sampling of explored regions. A simple rectilinear grid with equal bin sizes is used to uniformly bin the SPIB learned CVs. The undersampled regions are determined based on the expert-based bins by counting the number of samples within each bin. New WE simulations can then be performed by combining SPIB-based bins and expert-based bins. The SPIB and WE processes are iteratively applied until sufficient samples are obtained. If the simulations fail to explore new regions along the current direction, adjustments to the expert-based CVs and their associated bins can be made to explore other directions based on expertise. Finally, quantities of interest such as rates or free energy differences can be estimated from the samples.
    }
    \label{fig:flow-chart}
\end{figure}

\section{Method}
\label{sec:method}

\subsection{Weighted Ensemble (WE) Algorithm}
\label{sec:weighted_ensemble}

In a typical MD simulation, the system spends most of its time in high-probability basins, and rarely leaves the basins to visit low-probability regions such as transition and metastable states. The WE algorithm \cite{huber1996weighted,zuckerman2017weighted} adopts a splitting strategy using an ensemble of system copies (or “replicas”), each of which carries a statistical weight. These replicas are split or merged at a periodic time interval, called the resampling time $\tau$, guaranteeing an even coverage of the configuration space. Through this resampling procedure, the WE method is capable of enhancing the sampling of the low-probability regions. As these replicas are assigned statistical weights in a rigorous manner, in principle no bias is introduced in the dynamics. Thus, the WE algorithm can yield unbiased estimates of both the thermodynamic and kinetic properties of the molecular system. 

In WE, the resampling relies on dividing the configuration space into bins. Replicas within the same bin are replicated or pruned to maintain a target number of replicas per bin. Thus, these bins are at the heart of WE simulations, and the choice of the binning scheme can largely affect the performance of the WE algorithm. In the conventional pipeline, the configuration space will first be mapped to a small set of CVs, which is then divided into bins. However, selecting this CV set is challenging and often relies on expert intuition. In this paper, we will focus on how to construct this low dimensional CV space with the help of machine learning (ML), while still incorporating available expert intuition. Here, for simplicity, we adopt a uniform binning scheme, deferring the discussion of more advanced binning schemes to future investigations.

\subsection{State Predictive Information Bottleneck (SPIB)}
\label{sec:state_predictive_information_bottleneck}

State predictive information bottleneck (SPIB)\cite{Wang2021,wang2024information} uses an information bottleneck-based protocol to learn CVs of desired dimensionality from time resolved trajectory data, while simultaneously identifying important metastable states. Previous studies have demonstrated its efficacy in analyzing complex molecular systems by constructing multi-resolution MSMs.\cite{Wang2022,wang2024information} Furthermore, SPIB has proven valuable in learning low-dimensional CVs for enhanced sampling methods. This approach accelerates diverse processes, including ligand permeation\cite{Mehdi2022} and dissociation,\cite{Beyerle2024,lee2024calculating} protein conformational changes,\cite{Vani2023a,Vani2023b,gu2024empowering} and crystal polymorph nucleation.\cite{Zou2023,Wang2023}

SPIB adopts a structure akin to a variational autoencoder (VAE),\cite{kingma2013auto} consisting of two neural networks: An encoder maps the input features, such as those characterizing a protein’s current configuration, to a low-dimensional CV space, while a decoder tries to predict some target from this space. However, unlike VAE, where the aim is to reconstruct the entire input, SPIB only requires the learned CVs to predict coarse-grained future state labels, namely which state the system will occupy after a pre-specified lag time $\Delta t$. The main advantage of such a simplification of the prediction task is that only the motion related to the transitions between different states will be captured by the learned CVs, while the fluctuations inside any metastable state will be ignored. Typically, the number and location of such states are not available \textit{a priori}, but the SPIB method makes it possible to robustly estimate these on-the-fly.\cite{Wang2021,wang2024information} In this way, SPIB can autonomously cluster similar MD conformations into metastable states and accurately identify transition pathways between these states. This process allows SPIB to learn a low-dimensional CV space, where distinct states are separated into basins and connected by key transition pathways. Such clusterability of SPIB empowers the learned CVs to prioritize the representation of state-to-state transitions.

By introducing a lag time $\Delta t$, SPIB acts as a ``fast mode filter", selectively identifying metastable states with significant Boltzmann weights and lifetimes surpassing $\Delta t$, while excluding modes equilibrating faster.\cite{wang2024information} This enables SPIB to automatically pinpoint important metastable states in the system. $\Delta t$ serves as the minimum time resolution of interest. Aligning $\Delta t$ with the WE resampling time $\tau$—such as selecting $\Delta t$ as half of $\tau$—allows SPIB to filter out fast processes while retaining slower ones that cannot be sampled within $\tau$, making SPIB ideal for integration into WE simulations.

The CVs learned by SPIB can be interpreted as a low-dimensional continuous embedding of MD conformations that preserves maximal information about state-to-state transitions, effectively capturing all transitions between identified states.\cite{wang2024information} This is unlike commonly used dynamics-based dimension reduction methods such as tICA,\cite{Noe2013,Perez2013} which focus on learning a given number of slow processes. For instance, a 2D tICA captures only the two slowest processes while a 2D SPIB can accurately capture not just the two slowest, but all slow transition processes between the identified metastable states. This inherent distinction renders SPIB suitable for learning a low-dimensional CV space for systems featuring multiple metastable states of interest. 

Therefore, given these advantages, SPIB-learned CVs emerge as promising candidates for steering enhanced sampling methods. Leveraging these SPIB-learned CVs, we anticipate an efficient sampling process for WE simulations, even in the absence of specialized binning schemes.
 
\subsection{SPIB Augmented Enhanced Sampling}
\label{sec:spib_we}

The SPIB approach offers a promising way to automatically learn low-dimensional CVs from trajectory data. However, using data-driven methods to construct effective CVs requires a comprehensive sampling of the state-to-state transitions of interests in the system. This may seem impractical for enhanced sampling methods, such as weighted ensemble, commonly employed to investigate rare events occurring on long timescales and thus limited in initial stage sampling. This in fact presents a chicken-and-egg problem: finding reliable CVs requires collecting well-sampled state-to-state transitions, yet obtaining enough transitions relies on identifying suitable CVs to accelerate the sampling.

A solution to this conundrum could involve an iterative approach that alternates between rounds of enhanced sampling and SPIB to gradually improve sampling.\cite{Wang2019Machine, Mehdi2024} In the Reweighted Autoencoded Variational Bayes (RAVE) method,\cite{Ribeiro2018,Wang2019Past,Mehdi2022} the process starts with initial trajectories, and SPIB learns CVs, which may initially be suboptimal, to guide the subsequent enhanced sampling simulation to explore an increased amount of configuration space. Once a better sampling is collected, better CVs can then be learned. This looping of enhanced sampling and SPIB can continue until a satisfactory level of sampling is achieved.

However, as a deep neural network-based method, SPIB usually fails in extrapolation. This means that solely relying on SPIB to guide sampling may lead the system to get trapped in some metastable states, hindering effective exploration of new configurations. One solution to strengthen the extrapolability of such machine learned CVs is to use a linear encoder paired with a non-linear decoder in SPIB, as we have done in many previous applications.\cite{Beyerle2024,lee2024calculating,Vani2023a,Vani2023b,gu2024empowering,Zou2023,Wang2023} 

Here, we aim to explore an alternative approach by directly combining SPIB learned CVs with expert intuition based CVs (as illustrated in Fig. \ref{fig:illustration}). This is primarily motivated by recognizing that while experts may not grasp all the intricate dynamics of complex molecular systems, they possess intuition about crucial, more global demarcators of state-to-state transitions. Hence, expert-based CVs like root-mean-square deviation of atomic positions (RMSD), radius of gyration (Rg), or specific distances and dihedral angles, remain prevalent in many current enhanced sampling studies, despite the emergence of many ML-based CV identification approaches. These expert-based CVs excel in discerning states of interest, exhibit good extrapolation, and provide valuable insights into system dynamics. However, they may not capture all transition details accurately and can suffer from degeneracy, thus hindering the identification of important intermediate states and exact transition pathways for complex systems. Moreover, due to the prohibitive cost associated with sampling high-dimensional spaces, these expert-based CVs also are constrained by their dimensionality, posing additional challenges in proposing effective CVs. 

In contrast, ML-based CVs, such as those learned by SPIB, are inherently low-dimensional yet excel in discerning distinct conformations. They are adept at capturing dynamic details and identifying transition pathways, thereby significantly reducing the issue of degeneracy. However, despite their versatility and broad applicability, these methods often struggle with extrapolation. As a result, there is often a desire to incorporate system-specific intuition to guide exploration and CV identification processes in practical applications.

Hence, a simple yet effective approach is to combine the strengths of both classes of CVs, as we will elaborate on next.

\begin{figure}[htb]
    \centering
    \includegraphics[width=0.5\textwidth]{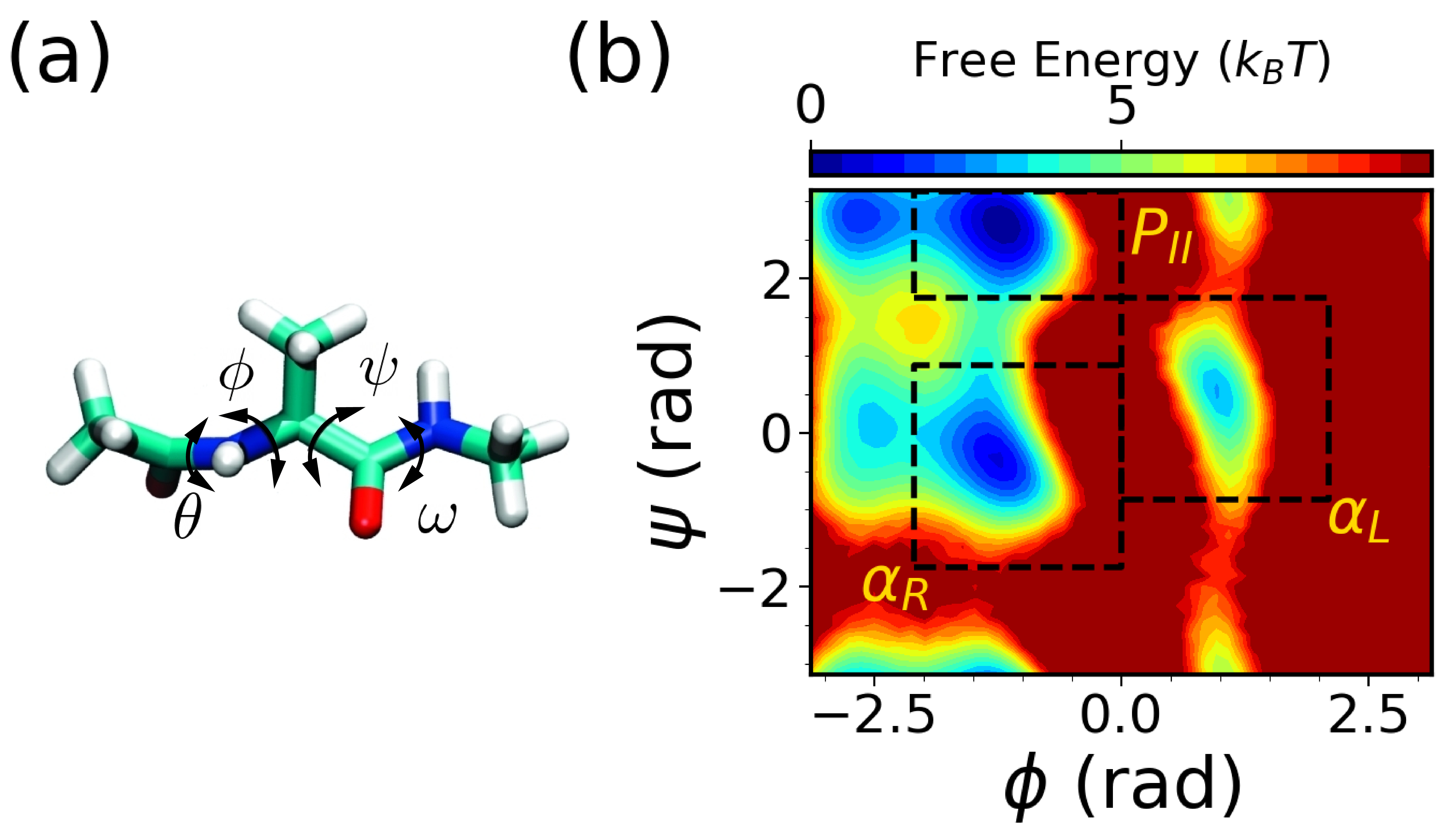}
    \caption{Characterizing the conformaion changes of alanine dipeptide in water. (a) Structure visulization and (b) the free energy surface obtained from the reference brute force simualtion along two dihedral angles: $\phi$ ($C$-$N$-$C_\alpha$-$C$), $\psi$ ($N$-$C_\alpha$-$C$-$N$). Rate constants to be calculated between the following regions of interest in alanine dipeptide: $\alpha_R$, $\alpha_L$ and $P_{II}$. These regions are described in boxes and defined as state $\alpha_R$: ($-120^{\circ}\le\phi\le 0^{\circ}$, $-100^{\circ}\le\psi\le 50^{\circ}$), $\alpha_L$: ($0^{\circ}\le\phi\le 120^{\circ}$, $-50^{\circ}\le\psi\le 100^{\circ}$) and $P_{II}$: ($-120^{\circ}\le\phi\le 0^{\circ}$, $100^{\circ}\le\psi\le 180^{\circ}$).}
    \label{fig:aladip_state_illustration}
\end{figure}

\begin{figure*}[htb]
    \centering
    \includegraphics[width=0.9\textwidth]{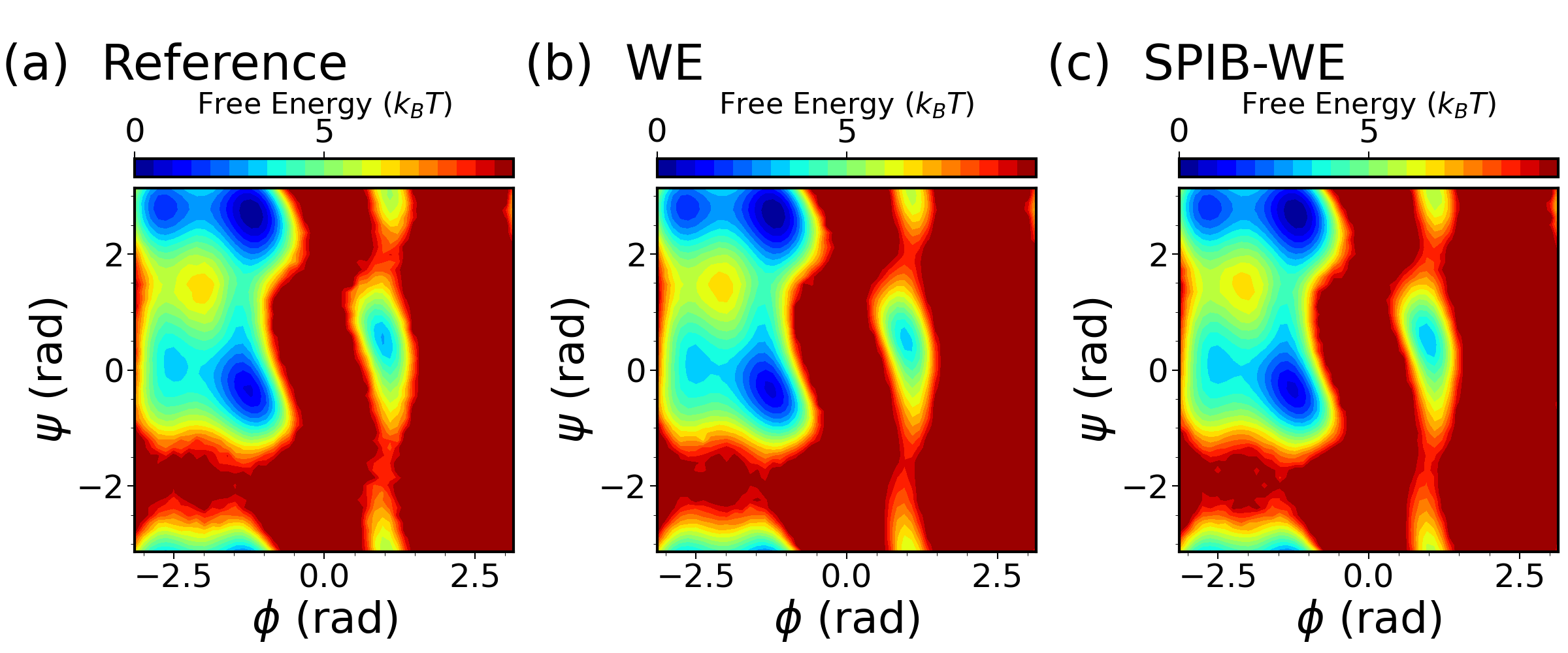}
    \caption{Average free energy surfaces of alanine dipeptide from three separate 10 $\mu s$ runs of (b) the WE solely based on expert-based CVs and (c) the hybrid SPIB-WE, compared with (a) the reference 4 $\mu s$ brute-force simulations.}
    \label{fig:aladip_FES_comparison}
\end{figure*}

\begin{figure*}[htb]
    \centering
    \includegraphics[width=0.9\textwidth]{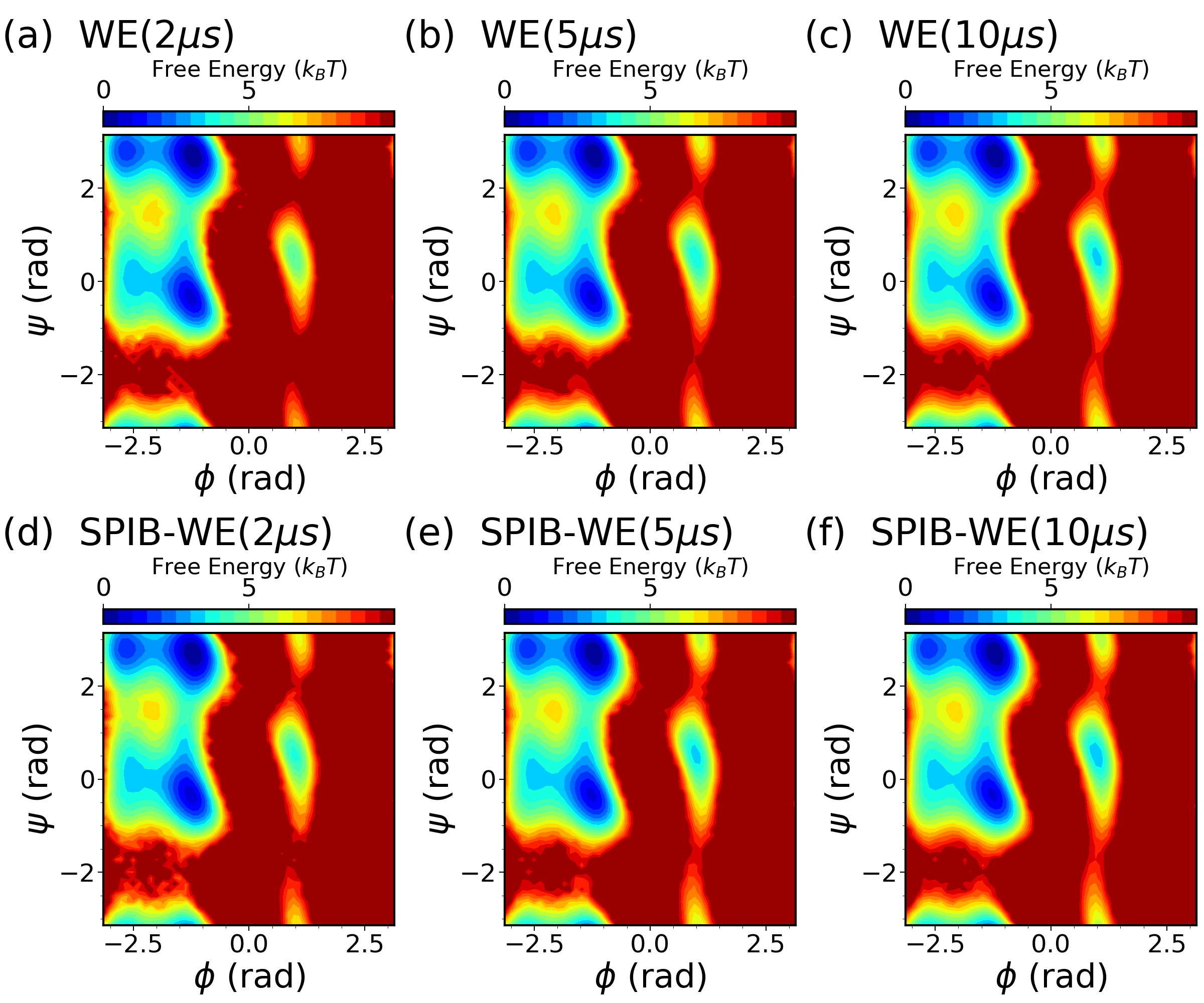}
    \caption{Average free energy surfaces of alanine dipeptide along $\phi$ and $\psi$ from three separate runs of (a-c) expert-based WE and (d-f) hybrid SPIB-WE at different stages of the aggregated simulations.}
    \label{fig:aladip_FES_evolution}
\end{figure*}

\subsection{A Hybrid Approach for CV construction for Weighted Ensemble}

In this subsection, we propose a simple strategy to combine expert-based CVs and data-driven SPIB-learned CVs for augmenting WE simulations. The idea is straightforward and broadly applicable, extending beyond just WE simulations: utilize deep learning based models to guide sampling in explored regions, while leveraging expert-based information to encourage exploration of new or undersampled regions. 

In this work, we adopt a uniform binning scheme for both expert-based CVs and SPIB-learned CVs. Therefore, combining expert-based and SPIB-learned CVs involves integrating bins derived from both sets of CVs. A simple strategy is proposed in Fig. \ref{fig:alg}. For instance, if a replica is assigned to bin $i_e$ based on the expert-based bins and to bin $i_s$ based on the SPIB-learned bins (where $i$ represents the bin index, and subscripts $e$ and $s$ denote expert-based and SPIB-learned, respectively), we need to determine when to use the expert-based binning and assign the replica to bin $i_e$ and when to use the SPIB-learned binning and assign the replica to bin $i_s$. First, we define the undersampled regions based on expert-based bins. Utilizing the trajectory data employed for SPIB training, we count the number of samples within each bin by mapping all samples to the expert-based bins, without considering sample weights. Subsequently, expert-based bins with counts below the average are identified as undersampled regions. This determination is updated after each SPIB training.
Then, we assess whether the replica is in undersampled regions. If a replica is in undersampled regions and there are fewer replicas mapped to bin $i_e$ than to bin $i_s$, the replica is assigned to bin $i_e$; otherwise, it is assigned to bin $i_s$. This hybrid strategy effectively combines the strengths of both SPIB-based CVs and expert-based CVs. In well-sampled regions, SPIB-learned CVs effectively separate metastable states and highlight key transition pathways, offering reliable guidance for enhanced sampling and facilitating slow state-to-state transitions. Conversely, in undersampled regions, the hybrid approach utilizes both SPIB-based bins and expert-based bins to improve sampling of undersampled transition states and promote exploration of new regions. This approach facilitates effective navigation of complex systems, ensuring efficient and robust sampling across the configuration space of interest.

The full workflow is summarized in Fig. \ref{fig:flow-chart}. Initially, utilizing expert-based CVs and their corresponding binning scheme, $N_0$ iterations of WE simulations are conducted to gather initial trajectories. Next, we train the SPIB model on the gathered MD data, clustering the sampled MD conformations into metastable states. This process allows SPIB to effectively map explored regions to a low-dimensional CV space, where distinct states are separated into basins connected by key transition pathways. A more detailed discussion of the neural network architecture and model training is provided in the Supporting Information (SI). In this work, we employ a rectilinear grid with equal bin sizes to uniformly bin the SPIB-learned CV space. We implement a straightforward algorithm to automatically adjust the bin size, aiming to maintain a roughly constant desired number of occupied SPIB bins (see SI for details). These SPIB-based bins, in combination with the expert-based bins, are then used to guide further WE simulations. $N_\text{update}$ iterations of WE simulations are conducted using these hybrid bins. Subsequently, based on the newly collected data, a new SPIB model can be obtained. This iterative process between WE simulations and SPIB training continues until a satisfactory level of sampling is achieved, allowing for the final estimation of quantities of interest, such as rates or free energy differences.

During the iterative process, all the metastable states and pathways of explored regions are effectively mapped into the SPIB-learned CVs. Thus, armed with these SPIB-learned CVs as a roadmap, adjustments can be made to the expert-based CVs and their associated bins whenever exploration in the current direction is impeded or when exploration of alternative directions is deemed necessary, as illustrated in the left half of Fig. \ref{fig:flow-chart}. This flexibility allows us to fully leverage expert intuition and numerous expert-based CVs, largely expanding the potential of this hybrid approach. While this may require more expertise and intervention, it provides users with greater control over the exploration process.

\subsection{Simulation Details}

To demonstrate our hybrid SPIB-WE approach, we benchmark it using two model systems: the small biomolecule alanine dipeptide in water and the mini folding chignolin mutant protein, CLN025, in implicit solvent. All simulations are performed with the OpenMM simulation package\cite{eastman2017openmm} patched with PLUMED 2.7\cite{Tribello2014} through the OpenMM-PLUMED interface plugin.\cite{Eastman2021} 

For alanine dipeptide, the PDB file was obtained from \href{http://ftp.imp.fu-berlin.de/pub/cmb-data/alanine-dipeptide-nowater.pdb}{http://ftp.imp.fu-berlin.de/pub/cmb-data/alanine-dipeptide-nowater.pdb}. The system was prepared using OpenMM with the AMBER ff14SB force field,\cite{maier2015ff14sb} and subjected to explicit solvation using the TIP3P water model. The MD integrations are performed with a timestep of 2\,fs. The system is first equilibrated under the constant number, volume, and temperature (NVT) ensemble for 10\,ns, followed by the equilibration in the constant number, pressure, and temperature (NPT) ensemble for 10\,ns. The resulting structure after this equilibration phase serves as the starting structure for all subsequent WE simulations. The NPT ensemble is finally employed for the WE simulations. We maintain a constant temperature of 300 K using the Langevin integrator with a collision frequency of 1 ps$^{-1}$, while the pressure is maintained at 1 bar using the Monte Carlo barostat.\cite{Aqvist2004}

We obtained the folded structure of CLN025 (amino acid sequence YYDPETGTWY) from the 2RVD entry of the Protein Data Bank. The system was prepared using OpenMM with the AMBER ff14SB-onlysc force field,\cite{maier2015ff14sb} and subjected to generalized Born (GB) implicit solvation using the GBn2 model.\cite{nguyen2013improved} The MD integrations are performed with a timestep of 2\,fs. The system is equilibrated under the NVT ensemble for 10 ns. The simulations are performed using the Langevin integrator at the melting temperature of 340 K with a collision frequency of 1 ps$^{-1}$. The resulting structure after equilibration is utilized as the starting structure for all WE simulations. The WE simulations for CLN025 are performed in the NVT ensemble. 

All WE simulations are carried out using the open-source WESTPA 2.0 software package.\cite{russo2022westpa} All simulation trajectories are saved with a time resolution of 1 ps for both systems, facilitating SPIB training, as well as rate and free energy profile estimation. The resampling time $\tau$ is set to 20 ps and the target number of replicas per bin is set to 4, identical for all WE and SPIB-WE simulations. We initially perform $N_0=20$ WE iterations for alanine dipeptide, and $N_0=40$ WE iterations for CLN025 before starting SPIB training. We use same SPIB update interval ($N_\text{update}=20$ WE iterations) for both systems. We observe that SPIB training time remains nearly unchanged as the training dataset size increases. Therefore, we train SPIB on data from the last $N_\text{last}=500$ WE iterations for both systems. We set the lag time of SPIB as the half of the WE resampling time, resulting in $\Delta t=10$ ps. For binning the SPIB-learned CVs, we set the desired number of occupied SPIB bins to 100 for both systems, ensuring a relatively fine grid along the 2D CVs while balancing computational costs.

We find that due to the relatively slow transition rates, some explored states in the initial stages of the SPIB-WE process may have very small weights. Considering these weights during SPIB training at the outset may lead SPIB to overlook these states. Typically, SPIB tends to disregard states with populations smaller than $10^{-3}$. Therefore, in this work, we ignore the weights during the first 100 WE iterations. After this initial period, all weights are considered as usual. As a general guideline for future users, it is recommended to ignore weights in the initial stages of the SPIB-WE protocol. If SPIB identifies too many states or if states of interest already have significant weights ($>10^{-3}$), then weights are recommended to be considered. This allows SPIB to ignore states with very small weights and focus on sampling transitions between more significant states.

\begin{table*}[htbp]
\caption{Rate Constants (in ns$^{-1}$) between Different Metastable States for alanine dipeptide. $^a$In the case of the brute-force simulation, the first value indicates the average rate constant, while the values within the brackets indicate the 95\% credible interval computed using Bayesian bootstrapping. $^b$For both the SPIB-WE and expert-based WE methods, the reported values represent the mean along with the standard error of the mean derived from three independent runs.}
\label{tab:aladip_rate}
 \begin{tabular}{|>{\centering\arraybackslash}p{3cm}| >{\centering\arraybackslash}p{4cm}| >{\centering\arraybackslash}p{4cm}| >{\centering\arraybackslash}p{4cm}|} 
 \hline
 transition & brute-force$^a$ & SPIB-WE$^b$ & WE$^b$ \\
 \hline
    $\alpha_R$ to P$_{II}$ & 8.39 [8.17, 8.63] & $8.26\pm0.05$ & $8.29\pm0.06$\\
    P$_{II}$ to $\alpha_R$ & 3.62 [3.52, 3.72] & $3.65\pm0.13$ & $3.590\pm0.011$\\
    $\alpha_R$ to $\alpha_L$ & 0.0113 [0.0090, 0.0145] & $0.0099\pm0.0018$ & $0.1030\pm0.0005$\\
    $\alpha_L$ to $\alpha_R$ & 0.38 [0.29, 0.49] & $0.43\pm0.06$ & $0.35\pm0.05$\\
    $\alpha_L$ to P$_{II}$ & 0.42 [0.32, 0.57] & $0.38\pm0.07$ & $0.339\pm0.013$\\
    P$_{II}$ to $\alpha_L$ & 0.0115 [0.0091, 0.0148] & $0.0099\pm0.0018$ & $0.0104\pm0.0006$\\
 \hline
\end{tabular}
\end{table*}

\begin{figure*}[htb]
    \centering
    \includegraphics[width=0.9\textwidth]{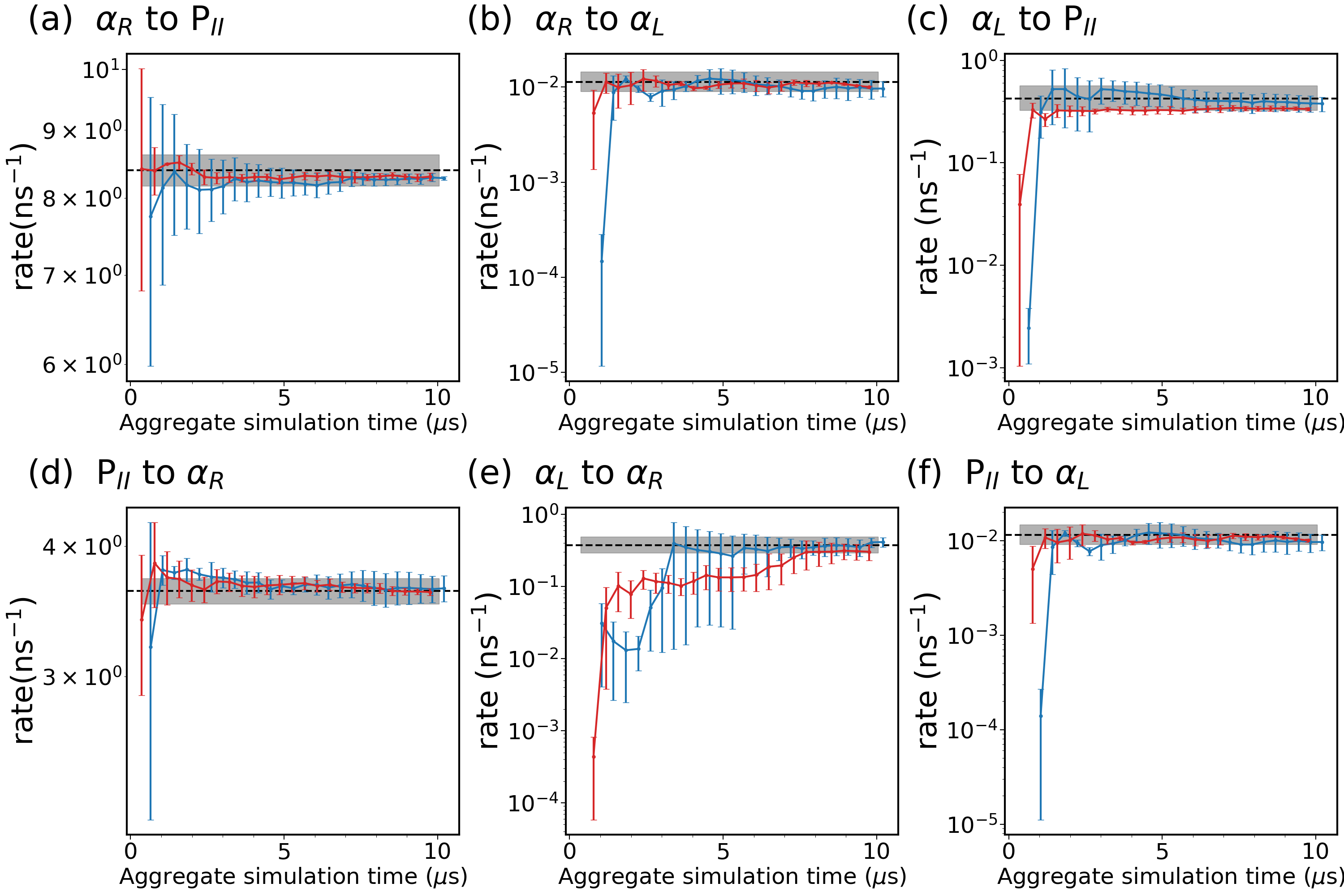}
    \caption{Evolution of rate constants over aggregate simulation time for the expert-based WE simulations (red), and the SPIB-WE approach (blue). The error bars represent the standard error of the mean from three independent runs. The black line shows the ground-truth obtained from the 4 $\mu$s brute-force simulation, and the shaded area shows the 95\% credible interval estimated using Bayesian bootstrapping.}
    \label{fig:aladip_rate_evolution}
\end{figure*}

\begin{figure}[htb]
    \centering
    \includegraphics[width=0.5\textwidth]{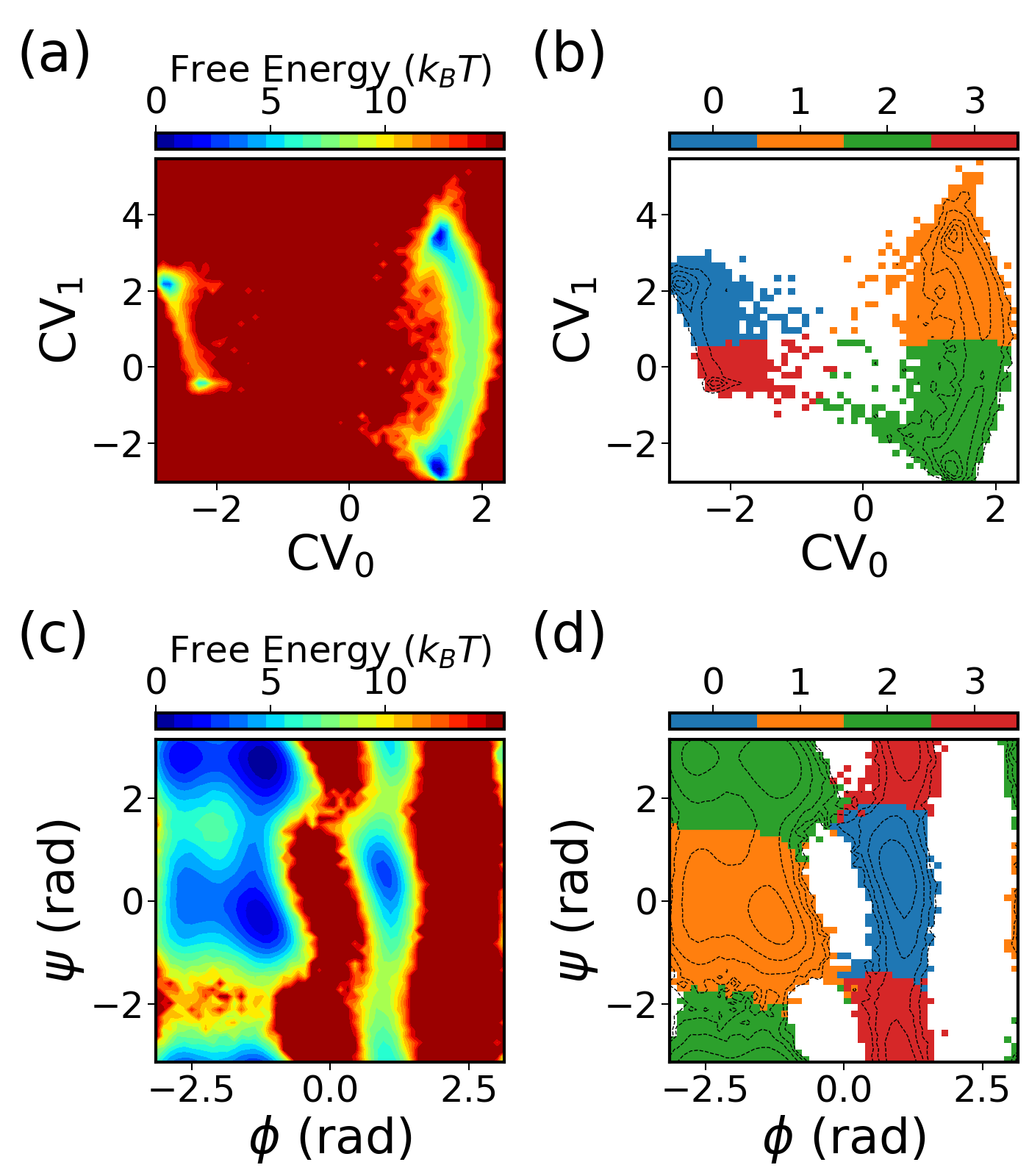}
    \caption{The CVs and metastable states learned by SPIB based on the last 500 WE iteration data for alanine dipeptide in water. The free-energy surface of the two-dimensional SPIB CVs is depicted in (a), while the learned metastable states in the 2D SPIB CVs are illustrated in (b). The free-energy surface of $\phi$ and $\psi$ is displayed in (c), while (d) presents the learned state labels projected onto the $\phi$-$\psi$ space.}
    \label{fig:aladip_spib_results}
\end{figure}

\begin{figure*}[htb]
    \centering
    \includegraphics[width=0.9\textwidth]{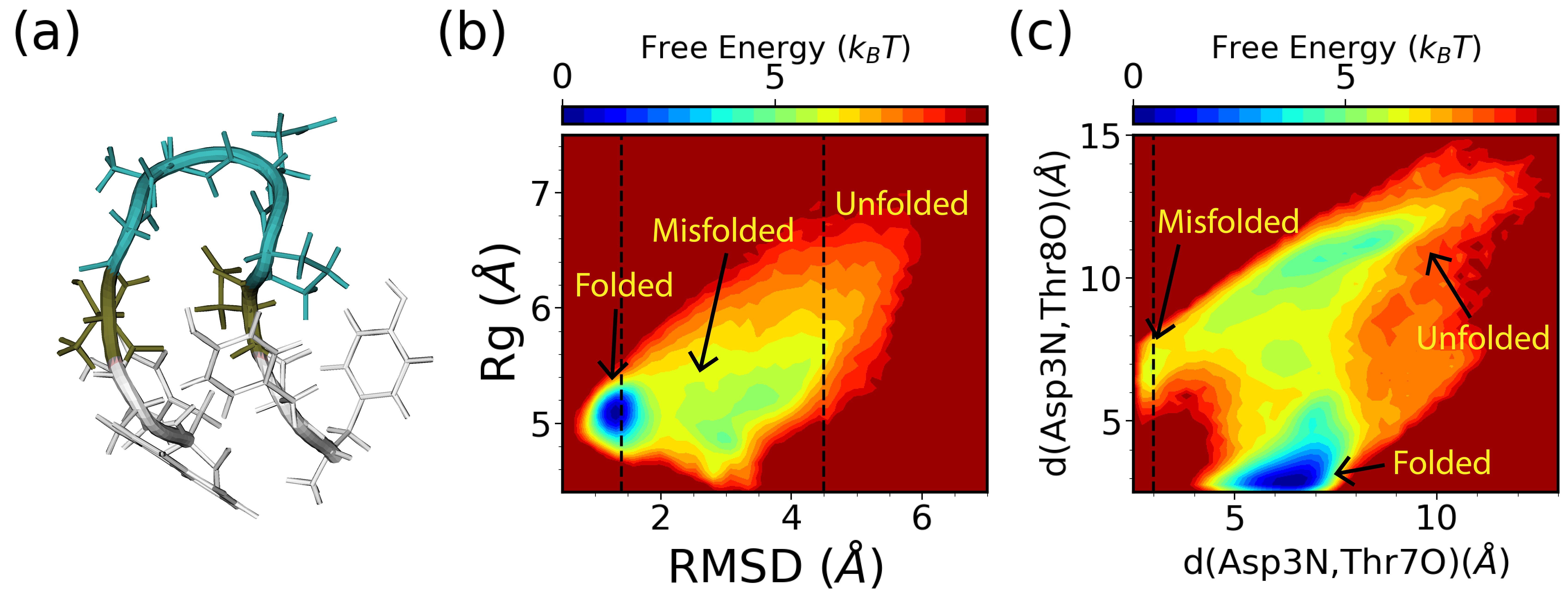}
    \caption{Characterizing the folding of CLN025 in implicit solvent. (a) Structure visulization (PDB ID:2RVD) and the free energy surfaces obtained from the reference brute force simualtions along (b) RMSD and Rg, and (c) $d(\text{Asp3N},\text{Tyr7O})$ and $d(\text{Asp3N},\text{Thr8O})$.}
    \label{fig:chignolin_state_illustration}
\end{figure*}

\begin{figure*}[htb]
    \centering
    \includegraphics[width=0.9\textwidth]{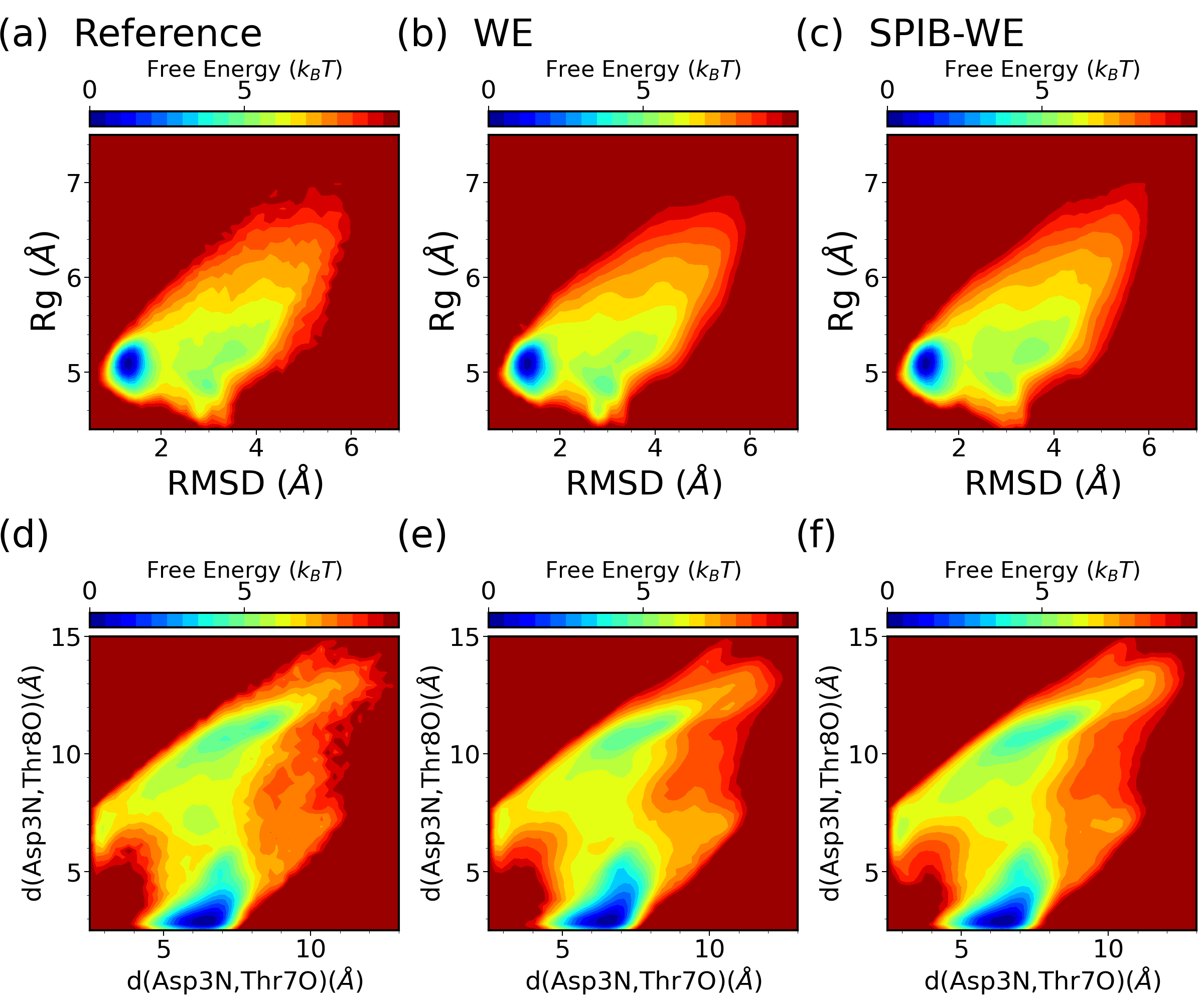}
    \caption{Average free energy surfaces of CLN025 from three separate 10 $\mu s$ runs of (b,e) the expert-based WE and (c,f) SPIB-WE, compared with the reference brute-force simulations (a,d).}
    \label{fig:chignolin_FES_comparison}
\end{figure*}

\begin{figure*}[htb]
    \centering
    \includegraphics[width=0.9\textwidth]{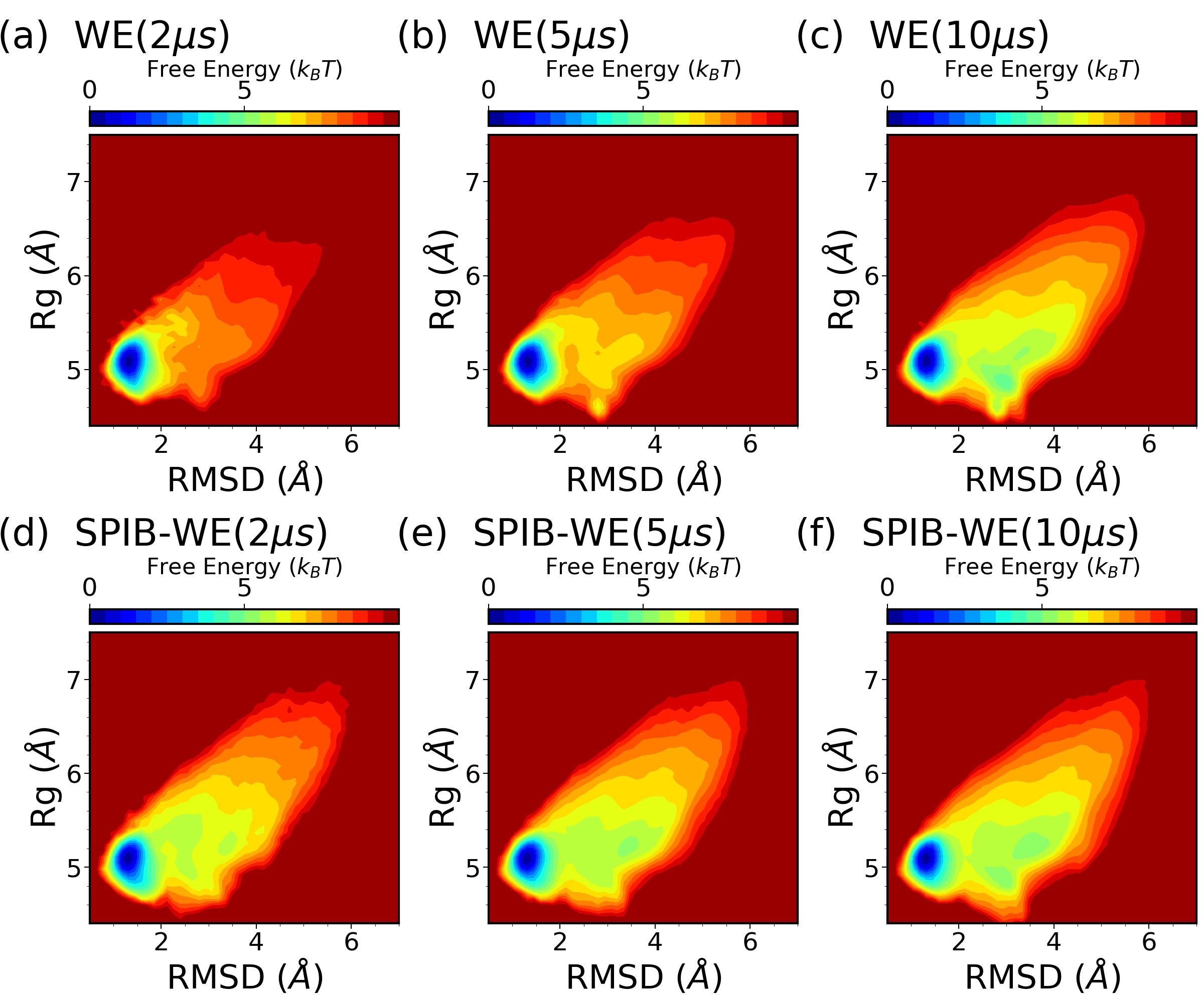}
    \caption{Average free energy surfaces of CLN025 along RMSD and Rg from three separate runs of (a-c) expert-based WE and (d-f) hybrid SPIB-WE at different stages of the aggregated simulations.}
    \label{fig:chignolin_FES_evolution_RMSD_Rg}
\end{figure*}

\section{Results}
\label{sec:Results}

\subsection{Alanine Dipeptide}

Our starting point is the very well-studied model system alanine dipeptide. It has 22 atoms with an acetyl group at the N-terminus and N-methylamide at the C-terminus. The backbone torsion angles $\phi$ and $\psi$ are known to be the most important CVs characterizing the dynamics of alanine dipeptide in water, as shown in Fig. \ref{fig:aladip_state_illustration}(a), which we use as the expert-based CVs for the WE simulations. Both CVs are evenly spaced in bins of 0.17 rad in the range of [-3.14 rad, 3.14 rad]. For SPIB training, pairwise distances between all heavy-atom serve as the SPIB input, resulting in 45 distances in total. The WE solely based on expert-based CVs is run for a total simulation time of 30 $\mu$s, averaged over three independent runs of 10 $\mu$s each. Similarly, the hybrid SPIB-WE approach is run for a total simulation time of approximately 30 $\mu$s, averaged over three independent runs of approximately 10 $\mu$s each. A brute-force 4 $\mu$s simulation is carried out to obtain reference values.

We first analyze the free energy surfaces (FES) of alanine dipeptide obtained from both the WE solely based on expert-based CVs and the hybrid SPIB-WE approaches in Fig. \ref{fig:aladip_FES_comparison}. The results show that both methods adequately sample the FES compared to the FES obtained from the reference long brute-force simulation. This underscores the capability of both the expert-based WE and the hybrid SPIB-WE methods to accurately capture the thermodynamics of the system. Fig. \ref{fig:aladip_FES_evolution} showcases the average FES from three separate runs of expert-based WE and hybrid SPIB-WE methods at different stages of the simulations.

We identify three metastable states in alanine dipeptide: $\alpha_R$, $\alpha_L$, and $P_{II}$, as defined in Fig. \ref{fig:aladip_state_illustration}(b). Subsequently, we assess both the expert-based WE and the SPIB-WE by estimating rate constants between these states, as summarized in Table \ref{tab:aladip_rate}. Fig. \ref{fig:aladip_rate_evolution} shows the evolution of the estimated rate constants over aggregate simulation time for the expert-based WE and SPIB-WE. Although the total simulation time to converge is similar to that of brute-force unbiased reference MD, it is important to note that WE runs simulations in parallel as a series of short trajectories. This parallelization with resampling significantly accelerates the convergence process, making WE more efficient despite the comparable total simulation time. In this case, since the expert-based CVs $\phi$ and $\psi$ effectively characterize the dynamics of alanine dipeptide in water, both the expert-based WE and the SPIB-WE accurately estimate the rate constants between the regions of interest. We note comparable convergence and accuracy of the kinetic rates for both methods, although the expert-based WE tends to exhibit smaller variance between different runs. These results serve as a validation that SPIB can learn CVs as effective as $\phi$ and $\psi$ from pairwise heavy-atom distances for the alanine dipeptide systems. For more complex systems, such as chignolin, where expert-based CVs struggle to capturing its dynamics, SPIB's advantage in managing intricate dynamics becomes evident, as illustrated in Figure 10.

Fig. \ref{fig:aladip_spib_results} exemplifies the SPIB learned CVs. Fig. \ref{fig:aladip_spib_results}(a) shows the FES of the SPIB learned CVs, while Fig. \ref{fig:aladip_spib_results}(b) exhibits the metastable states learned by the algorithm. Utilizing a lag time of $\Delta t=10$ ps, SPIB effectively discerns four metastable states from the data, aligning with the well-known free energy minima in the $\phi$-$\psi$ space illustrated in Fig. \ref{fig:aladip_spib_results}(c,d). Notably, the SPIB-learned CVs primarily highlight state-to-state transitions, while fluctuations within each state are predominantly mapped to a single point. This makes it efficient in encouraging sampling even with uniform binning.

\subsection{Chignolin mutant CLN025}

The chignolin mutant, CLN025, with a sequence of YYDPETGTWY is one of the smallest fast-folding proteins\cite{Honda2008} that folds into a $\beta$-hairpin as shown in Fig. \ref{fig:chignolin_state_illustration}(a). For this system, the commonly expert-based CVs for WE simulations are mass-weighted RMSD to the folded structure (PDB ID:2RVD) and mass-weighted Rg.\cite{ojha2023deepwest,ahn2021gaussian} We place both CVs evenly in bins of 0.03 nm in the range of [0.04 nm, 0.70 nm] for RMSD and in the range of [0.41 nm, 0.83 nm] for Rg. For SPIB training, nearest-neighbor heavy-atom distances between all residues separated by two or more residues serve as the SPIB input, resulting in 28 distances in total. The WE solely based on these two expert-based CVs is executed for three independent runs of 10 $\mu$s each. Similarly, the hybrid SPIB-WE approach is conducted for three independent runs of approximately 10 $\mu$s each. Additionally, three brute-force reference unbiased 4 $\mu$s simulations were carried out to obtain reference values.

To compare the performance of the hybrid SPIB-WE method with the regular WE methods, we focus on three metastable states: the folded state, unfolded state, and misfolded state in CLN025. The folded region is defined as RMSD $<$ 0.14 nm, while the unfolded region is defined as RMSD $>$ 0.45 nm. Through brute-force simulations, we observe that even at its melting temperature (340 K), the folded state remains the most dominant state in implicit solvent. To better define the misfolded state, we introduce another two CVs, the hydrogen bond distance between Asp3O and Tyr7O and the hydrogen bond distance between Asp3N and Tyr8O. Thus, the misfolded state is defined as $d(\text{Asp3N},\text{Tyr7O}) < 0.3$ nm, as shown in Fig. \ref{fig:chignolin_state_illustration}(b,c). Note that these two hydrogen bond distance CVs are only used for defining the states and examining the FES. Indeed, despite its apparent instability, we observe sampling of the misfolded state in the brute-force simulations.

We first analyze the FES of CLN025 from the expert-based WE and the hybrid SPIB-WE approaches in Fig. \ref{fig:chignolin_FES_comparison}. Once again, the results show that both methods are able to sample the folded, unfolded and misfolded states, resulting in reasonably well FES along the selected CVs, compared to the FES obtained from the reference long brute-force simulations. This demonstrates that both methods can sample the thermodynamics of the system accurately. Fig. \ref{fig:chignolin_FES_evolution_RMSD_Rg} and Fig. S1 demonstrate the average FES from three separate runs of expert-based WE and SPIB-WE at different stages of the simulations. We find that SPIB-WE appears to achieve faster convergence of the average FES in the unfolded and mifolded regions where $RMSD > 2$\AA.

\begin{table*}[htbp]
\caption{Rate Constants (in ns$^{-1}$) between different metastable states for CLN025. $^a$In the case of the brute-force simulations, the first value indicates the average rate constant, while the values within the brackets indicate the 95\% credible interval computed using Bayesian bootstrapping. $^b$For both the SPIB-WE and expert-based WE methods, the reported values represent the mean along with the standard error of the mean derived from three independent runs.}
\label{tab:chignolin_rate}
 \begin{tabular}{|>{\centering\arraybackslash}p{4cm}| >{\centering\arraybackslash}p{4cm}| >{\centering\arraybackslash}p{4cm}| >{\centering\arraybackslash}p{4cm}|} 
 \hline
 transition & brute-force$^a$ & SPIB-WE$^b$ & WE$^b$ \\
 \hline
    folded to unfolded & 0.018 [0.013, 0.024] & $0.028\pm 0.004$ & $0.026\pm 0.010$\\
    unfolded to folded & 0.079 [0.046, 0.127] & $0.125\pm 0.022$ & $0.095\pm 0.034$\\
    folded to misfolded & 0.020 [0.015, 0.026] & $0.030\pm 0.007$ & $0.024\pm 0.009$\\
    misfolded to folded & 0.11 [0.06, 0.19] & $0.180\pm 0.026$ & $0.124\pm 0.045$\\
    misfolded to unfolded & 0.074 [0.047, 0.126] & $0.084\pm 0.011$ & $0.073\pm 0.028$\\
    unfolded to misfolded & 0.051 [0.037, 0.073] & $0.056\pm 0.011$ & $0.039\pm 0.015$\\
 \hline
\end{tabular}
\end{table*}

\begin{figure*}[ht]
    \centering
    \includegraphics[width=0.9\textwidth]{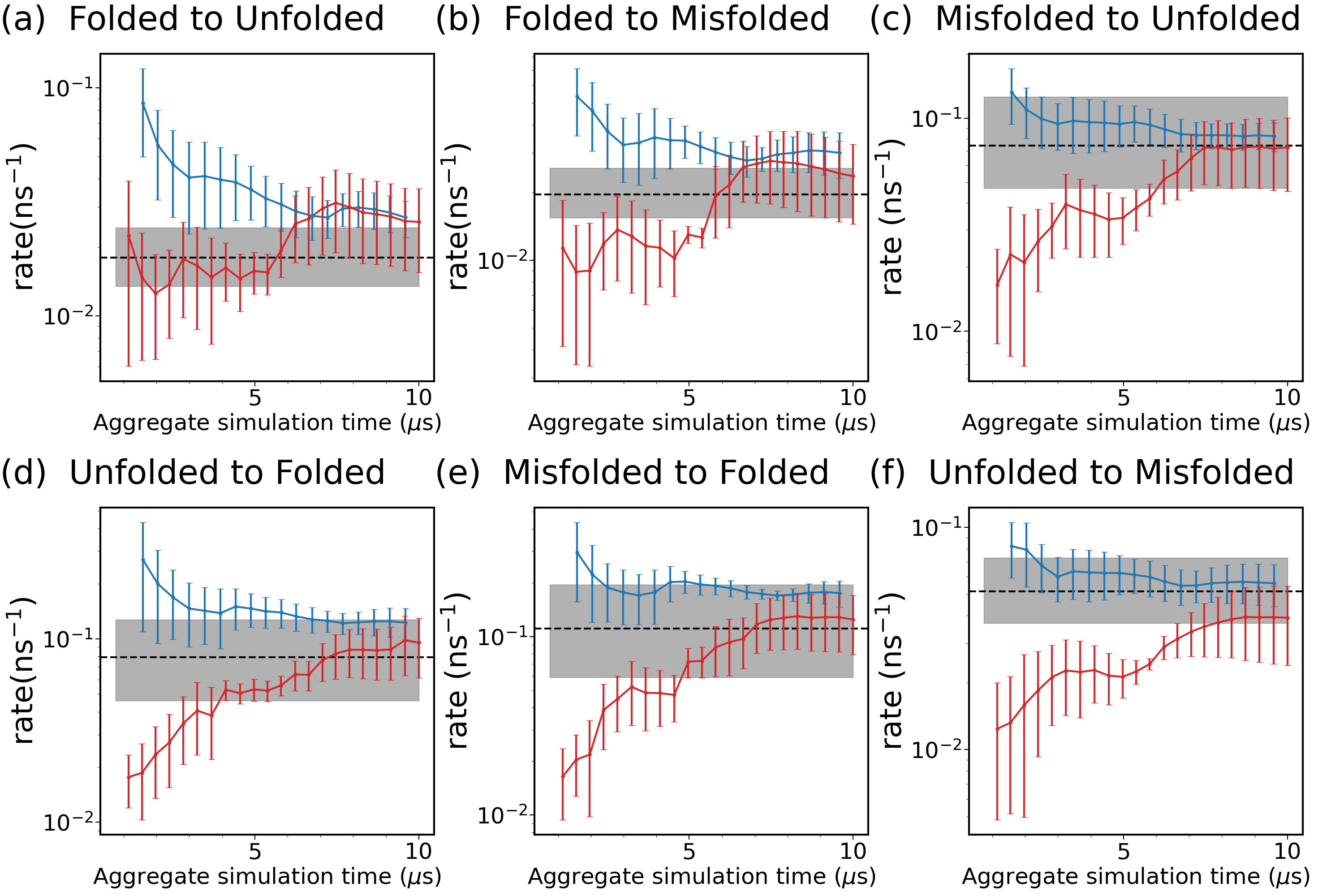}
    \caption{Evolution of rate constants over aggregate simulation time for the WE simulations (red), and the SPIB-WE approach (blue). For the WE and SPIB-WE methods, the error bars represent the standard error of the mean from three independent runs. The black line shows the ground-truth obtained from the long brute-force simulations, and the shaded area shows the 95\% credible interval estimated from 3 independent runs. The first 1 $\mu$s simulation is excluded from the rate analysis.}
    \label{fig:chignolin_rate_evolution}
\end{figure*}

We then test the SPIB-WE method by estimating rate constants between regions of interest, which are summarized in Table \ref{tab:chignolin_rate}. Fig. \ref{fig:chignolin_rate_evolution} shows the evolution of the estimated rate constants over aggregate simulation time for the expert-based WE and the hybrid SPIB-WE. We find that SPIB-WE generally exhibits a relative faster convergence compared with the regular WE method, with a smaller standard error of the mean at the same aggregate simulation time. Notably, while the expert-based CVs RMSD and Rg effectively sample the unfolding process, resulting in comparable performance with SPIB-WE in estimating rates for transitions from the folded state to unfolded or misfolded states (Fig. \ref{fig:chignolin_rate_evolution}(a,b)), they are less effective at sampling the folding process. This leads to slower convergence in estimating rates for other state-to-state transitions (Fig. \ref{fig:chignolin_rate_evolution}(c-f)), even though these transitions are actually much faster than the unfolding process and should be easier to sample.

Since no bias is introduced in either the expert-based WE or SPIB-WE, we attribute SPIB-WE's superior performance to its enhanced capability to distinguish between different metastable configurations and pinpoint crucial state-to-state transitions. This increased discriminatory power leads to a higher number of uncorrelated successful transition events, thereby enhancing SPIB-WE's effectiveness over expert-based WE in capturing state-to-state transitions and reducing correlations arising from resampling processes. This, in turn, may explain the reduction in run-to-run variances of rate esimation and the faster convergence of the FES and rates from SPIB-WE. 

\begin{figure}[ht]
    \centering
    \includegraphics[width=0.5\textwidth]{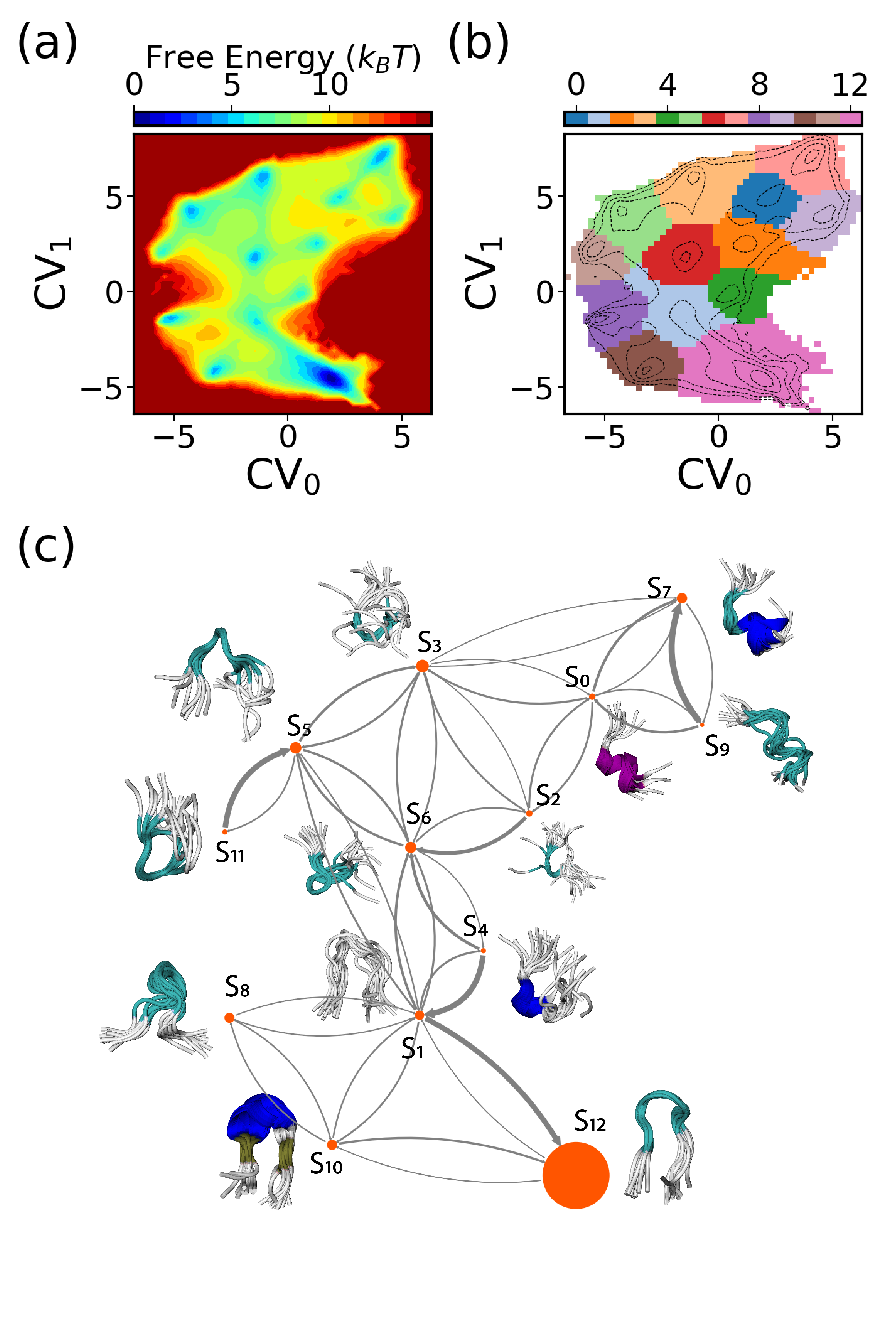}
    \caption{The CVs and metastable states learned by SPIB for CLN025. The free-energy surface of the two-dimensional SPIB CVs is depicted in (a), while the learned metastable states in the 2D SPIB CVs are illustrated in (b). (c) The MSM constructed based on states identified by SPIB is visualized using a flux network. The node size is proportional to the stationary population of the states, and the arrow width is scaled according to jump probabilities. Additionally, ten randomly selected conformations from each state are overlapped and displayed adjacent to the corresponding node.}
    \label{fig:chignolin_spib_results}
\end{figure}

For a more in-depth analysis of the SPIB-learned CVs, we aggregate data from the last 500 WE iterations of three independent SPIB-WE runs and retrain a single SPIB model, as depicted in Fig. \ref{fig:chignolin_spib_results}. Fig. \ref{fig:chignolin_spib_results}(a) showcases the free energy surface (FES) of the learned CVs, while Fig. \ref{fig:chignolin_spib_results}(b) illustrates the metastable states identified by the algorithm. Utilizing a lag time $\Delta t=10$ ps, we identify 13 metastable states from the data. The conformational ensembles and associated transitions for these 13 macrostates are visually represented in Fig. \ref{fig:chignolin_spib_results}(c). S${10}$ and S${12}$ correspond to the folded state, while S${2}$, S${3}$, and S${6}$ represent the collapsed unfolded states. We also observe a misfolded-like conformation in S${5}$, where CLN025 adopts an $\alpha$-turn from Asp3 to Gly7 instead of the native $\pi$-turn from Asp3 to Thr8. However, as suggested by the FES in Fig. \ref{fig:chignolin_FES_comparison}, this misfolded conformation is not very stable in implicit solvent simulations, consistent with findings from the previous study.\cite{chen2021machine} Interestingly, our SPIB-WE protocol is also able to sample conformations where CLN025 forms a helix structure, as seen in S$_{0}$, S$_{7}$, and S$_{9}$. As all these identified states have significant Boltzmann weights, we indeed observe that they are also sampled by expert-based WE. However, since the FES of the expert-based CVs shows no distinct basins, it would be more challenging to sample and identify these states using expert-based CVs. This result showcases how SPIB automatically clusters sampled MD conformations into important metastable states and maps all key state-to-state transitions into a low-dimensional CV space. Therefore, by utilizing these learned CVs to guide sampling, the WE simulations can better focus on important state-to-state transitions, thus achieving improved performance.

\section{Discussion}

In this work, we have proposed a hybrid approach to enhance weighted ensemble (WE) simulations that facilitates combining human expert based knowledge with deep learning derived descriptors. Our framework makes it possible to use deep learning to characterize well-explored regions of configuration space, but use human expertise to facilitate extrapolation into new regions. This circumvents the well known limitation of deep neural networks at extrapolating beyond the training set. Specifically, we use the State Predictive Information Bottleneck (SPIB) method to construct low-dimensional CVs which facilitate more efficient sampling of slow state-to-state transitions. This is particularly advantageous for guiding WE simulations, especially in systems characterized by multiple metastable states of interest. SPIB achieves this by automatically clustering MD conformations into metastable states based on a controllable lag time parameter and encoding all transitions between these metastable states into the learned CVs. Acknowledging the value of expert intuition in further enhancing the sampling of complex molecular systems, we complement SPIB-learned CVs with expert-based CVs. Here, SPIB-learned CVs bolster sampling in explored regions, while expert-based CVs steer exploration in regions of interest. This hybrid strategy harnesses the advantages of both expert-based and data-driven CVs, facilitating more efficient and reliable navigation through the complex configuration space. 

We validate our hybrid algorithm using two model systems, alanine dipeptide and chignolin mutant CLN025, demonstrating its efficacy in guiding the sampling of states of interest. While in cases like alanine dipeptide where expert-based CVs suffice, the hybrid approach only shows comparable performance to the solely expert-based approach. However, in more practical scenarios like CLN025, where perfect expert-based CVs are unavailable, the hybrid approach outperforms, showing faster convergence of free energy surfaces and rates with reduced run-to-run variance. Furthermore, we find SPIB is able to learn more important metastable states, which are clearly separated as basins in the FES of the learned low-dimensional CVs. This clear separation makes sampling and identifying these states much easier using SPIB-learned CVs compared to the expert-based CVs. Finally, the learned SPIB model offers insightful 2D visualizations of the dynamics, identifying crucial metastable states and pathways, thus providing a promising avenue for enhancing the analysis of the WE data.

In this work, we exclusively tested our method on well-studied model systems. However, we anticipate that using more expert-based CVs will be beneficial for tackling more challenging systems. In such cases, users can adjust the expert-based CVs whenever exploration within the current set is hindered. While this approach may require more manual intervention, it provides users with greater flexibility, leveraging all available expert-based CVs to guide exploration effectively. Moreover, we employ a rectilinear grid with equal bin sizes for binning both the expert-based and SPIB-learned CVs for simplicity, but more advanced binning schemes could further enhance sampling efficiency.\cite{torrillo2021minimal} Additionally, to accelerate the convergence of WE simulations, it is also promising to initialize the WE simulations with better weighted initial configurations, as demonstrated in previous studies.\cite{ojha2023deepwest,ahn2021gaussian} These initial configurations can be selected by training a SPIB on short unbiased or biased trajectories, similar to the approach used in Ref. \onlinecite{ojha2023deepwest}. Finally, considering SPIB's utility in constructing MSMs,\cite{wang2024information} another natural way to speed up the convergence of rate and FES estimation is through employing more efficient analysis tools like MSM and haMSM to analyze the WE data. We leave the exploration of these extension directions for future work.
\newline

\textbf{Supplementary material\newline }
See supplementary material for further numerical details. \newline

\textbf{Acknowledgements\newline }
This work was supported by NIH/NIGMS under award number R35GM142719 (P.T.). We thank UMD HPC’s Zaratan and NSF ACCESS (project CHE180027P) for computational resources. P.T. is an investigator at the University of Maryland-Institute for Health Computing, which is supported by funding from Montgomery County, Maryland and The University of Maryland Strategic Partnership: MPowering the State, a formal collaboration between the University of Maryland, College Park and the University of Maryland, Baltimore. \newline

\textbf{Code availability statement\newline }
The SPIB-WE package is available for public use at \href{https://github.com/wangdedi1997/spib_we}{https://github.com/wangdedi1997/spib\_we}. \newline

\textbf{References}
\bibliography{references}

\begin{figure*}[b!]
    \centering
    \includegraphics[width=0.8\textwidth]{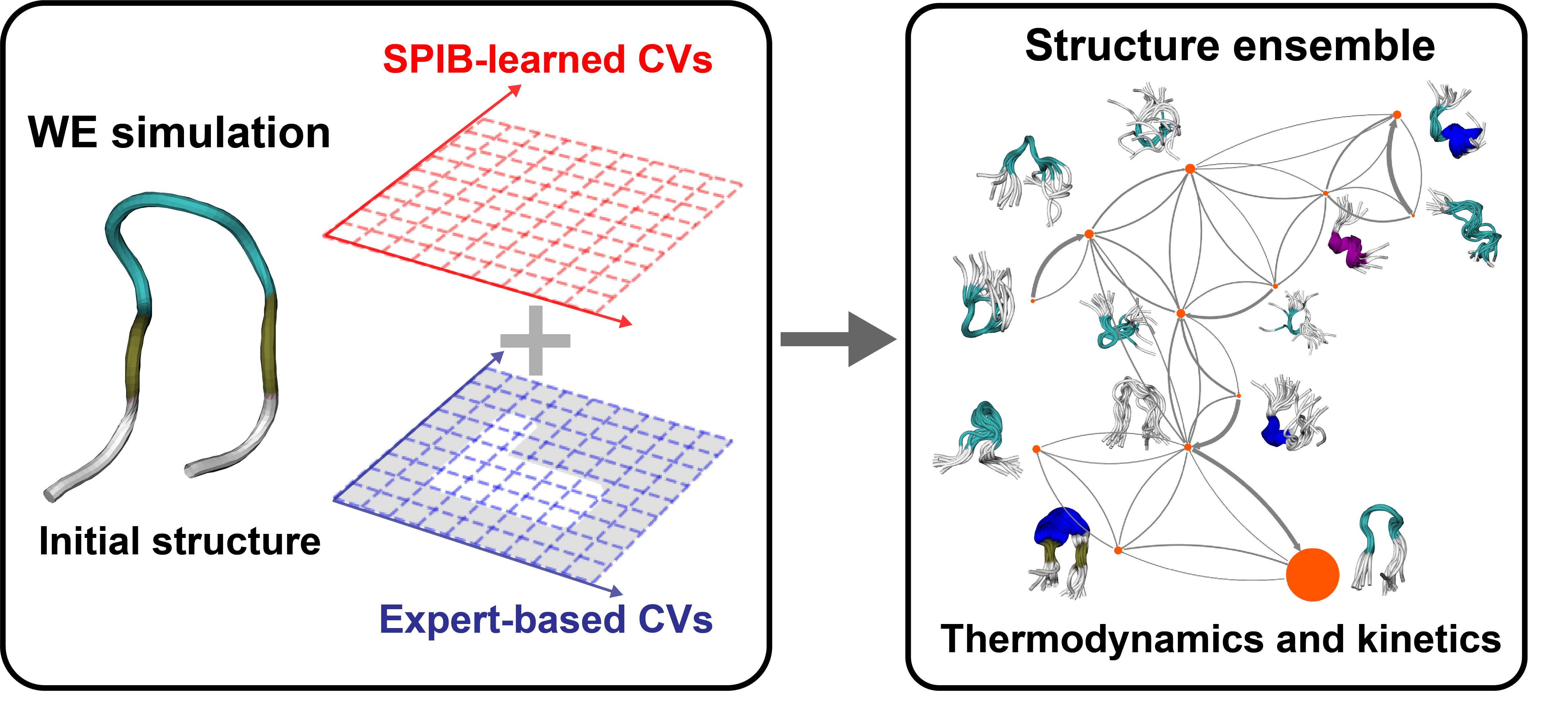}
    \caption*{Table of Contents graphic}
    \label{fig:toc}
\end{figure*}

\end{document}